\documentclass[preprint,aps,amssymb,showpacs,superscriptaddress,nofootinbib]{revtex4}
\usepackage{graphicx}

\newcommand{\Case}[2]{{\textstyle \frac{#1}{#2}}}

\begin{document}

\preprint{IMSc/2006/6/17}

\title{On obtaining classical mechanics from quantum mechanics}

\author{Ghanashyam Date}
\email{shyam@imsc.res.in}
\affiliation{The Institute of Mathematical Sciences,
CIT Campus, Chennai-600 113, INDIA}

\pacs{04.60.Pp,98.80.Jk,98.80.Bp}

\begin{abstract}

Constructing a classical mechanical system associated with a given
quantum mechanical one, entails construction of a classical phase space
and a corresponding Hamiltonian function from the available quantum
structures and a notion of coarser observations.  The Hilbert space of
any quantum mechanical system naturally has the structure of an infinite
dimensional symplectic manifold (`quantum phase space'). There is also a
systematic, quotienting procedure which imparts a bundle structure to
the quantum phase space and extracts a classical phase space as the base
space. This works straight forwardly when the Hilbert space carries
weakly continuous representation of the Heisenberg group and one
recovers the linear classical phase space $\mathbb{R}^{\mathrm{2N}}$. We
report on how the procedure also allows extraction of non-linear
classical phase spaces and illustrate it for Hilbert spaces being finite
dimensional (spin-j systems), infinite dimensional but separable
(particle on a circle) and infinite dimensional but non-separable
(Polymer quantization). To construct a corresponding classical dynamics,
one needs to choose a suitable section and identify an effective
Hamiltonian.  The effective dynamics mirrors the quantum dynamics
provided the section satisfies conditions of semiclassicality and
tangentiality. 

\end{abstract}


\maketitle

Developing a semi-classical approximation to quantum dynamics is in
general a non-trivial task. Intuitively, such an approximation entails
an {\em adequate} class of observable quantities (eg. expectation values
of self-adjoint operators) whose time evolution, dictated by quantum
dynamics, is {\em well approximated} by {\em a classical Hamiltonian
evolution}. Roughly, the adequate class refers to (say) basic functions
on a classical phase space (symplectic manifold) with a Hamiltonian
which is a function of these basic functions. The accuracy of an
approximation is controlled by how well the classically evolved
observables stay close to the quantum evolved ones within a given
precision specified in terms of bounds on quantum uncertainties. Having
a description of the quantum framework as similar as possible to a
classical framework is obviously an aid in developing semi-classical
approximations. 

Such a description is indeed available and is referred to as geometrical
formulation of quantum mechanics \cite{AshtekarSchilling}. The quantum
mechanical state space, a projective Hilbert space, is naturally a
symplectic manifold, usually infinite dimensional (finite dimensional
for spin systems). Furthermore, dynamics specified by a Schrodinger
equation is a Hamiltonian evolution. This is true for all quantum
mechanical systems. In addition, there is also a systematic quotienting
procedure to construct an associated Hamiltonian system (usually of
lower and mostly finite dimensions) which views the quantum state space
as a bundle with the classical phase space as its base space. This works
elegantly when the quantum Hilbert space is obtained as the weakly
continuous representation of a Heisenberg group.  Generically these are
separable Hilbert spaces and the extracted classical phase spaces are
{\em linear}, $\mathbb{R}^{\mathrm{2 n}}$.  

Quantum mechanical Hilbert spaces however arise in many different
ways.  For example, the (kinematical) Hilbert space of loop quantum
cosmology carries a {\em non-weakly continuous} representation of the
Heisenberg group and is non-separable.  For examples such as particle
on a circle and spin systems, one does not even have the Heisenberg
group. A semi-classical approximation is still needed for such
systems. Likewise, in classical mechanics (even for finitely many
degrees of freedom), the classical phase space is not necessarily
{\em linear} (eg the cylinder for particle on a circle, reduced phase
spaces of constrained systems etc). It is important to develop a
quotienting procedure to construct such, possibly non-linear,
classical phase spaces from more general quantum state spaces.  In
this work we develop such a procedure and illustrate it for three
examples: arbitrary spin-J system, particle on a circle and Bohr or
polymer quantization appearing in loop quantum cosmology (LQC).  This
takes care of the kinematical aspects. 

To construct an associated classical dynamics one has also to obtain
a Hamiltonian function (an {\em effective Hamiltonian}) on the
classical phase space. This is done by choosing a section of the
bundle and obtaining the effective Hamiltonian on the base space as a
pull back of the quantum mechanically defined one. An effective
Hamiltonian so defined, depends on the section chosen. One can now
construct two trajectories on the classical phase space: (a)
projection of a {\em quantum trajectory} (i.e. trajectory in the
quantum state space) onto the base space and (b) a trajectory in the
base space, generated by the effective Hamiltonian function. In
general, i.e.  for arbitrary sections, these two trajectories do {\em
not} coincide.  They do so when the section is tangential to the
quantum trajectories (equivalently when the section is preserved by
quantum dynamics). Since the classical states are obtained from
expectation values (via the quotienting procedure), for the classical
trajectories to reflect the quantum one, {\em within a certain
approximation}, it is necessary that the quantum uncertainties also
remain bounded within {\em prescribed tolerances}. In other words,
the states in the section should also satisfy conditions of
semiclassicality. 

In section I, we recall the basic details of the geometric formulation
from \cite{AshtekarSchilling} and describe the quotienting procedure in
a general setting. In general the classical phase is obtained as a {\em
sub-manifold} of the base space of the bundle. The general procedure is
then illustrated with three examples in the three subsections. 

Section II contains a discussion of dynamics and the conditions for
developing semi-classical approximation.  The classical phase space
being not the same as the base space in general, puts a first
requirement that the projection of sufficiently many quantum
trajectories onto the base space should actually be confined to the
sub-manifold of classical phase space. For constructing a Hamiltonian
dynamics on the classical phase space, a section over the classical
phase space needs to be chosen which provides an embedding of the
classical phase space into the quantum state space.  For having a useful
semiclassical approximation the section has to satisfy two main
conditions of semiclassicality and tangentiality.  The state space of
polymer quantization brings out further points to pay attention to while
developing a semi-classical approximation. 

In the last section we conclude with a summary and a discussions.

\section{Kinematical set-up} 
\label{Introduction} 
This section is divided in five subsections. The first one recalls the
symplectic geometry of the quantum state space.  This is included for a
self contained reading as well as for fixing the notation, experts may
safely skip this section.  More details are available in
\cite{AshtekarSchilling,Schilling}.  The second one describes the
quotienting procedure in some generality (but still restricted to
finitely many classical degrees of freedom) to get candidate classical
phases spaces which could also be non-linear. The next three
sub-sections illustrate the quotienting procedure for the examples of
spin-j system, particle on a circle and isotropic, vacuum loop quantum
cosmology.

\subsection{Symplectic geometry of (projective) Hilbert space}
\label{SymGeomRev}

Let $\cal{H}$ be a complex Hilbert space, possibly non-separable and let
$\cal{P}$ be the corresponding projective Hilbert space: the set of
equivalence classes of non-zero vectors of the Hilbert space modulo
scaling by non-zero complex numbers. Equivalently, if $\cal{S}$ denotes
the subset of normalized vectors, then $\cal{P} = \cal{S}/\mathrm{phase
~ equivalence}$. Unless stated otherwise, the Hilbert space is assumed
to be infinite dimensional.

Any complex vector space can be viewed as a real vector space with an
almost complex structure defined by a bounded linear operator $J, \ J^2
= -\mathbb{I}$ defined on it and multiplication by a complex number $a +
i b$ being represented as $a + b J$. The Hermitian inner product of the
complex Hilbert space can then be expressed in terms of a symmetric and
an anti-symmetric quadratic forms, $G(\cdot, \cdot)$ and $\Omega(\cdot,
\cdot)$ respectively. Explicitly,
\begin{equation}
\langle\Psi, \Phi\rangle ~ := ~ \Case{1}{2 \hbar} G(\Psi, \Phi) +
\Case{i}{2 \hbar} \Omega(\Psi, \Phi).
\end{equation}
The non-degeneracy of the Hermitian inner product then implies that $G$
and $\Omega$ are (strongly) non-degenerate. The real part $G$ will play
no role in this paper.

For a separable Hilbert space, let $|e_n\rangle$ denote an orthonormal
basis so that we have $|\Psi\rangle = \sum_n \psi^n |e_n\rangle, \
\psi^n \in \mathbb{C}$. Writing $\psi^n := x^n + i y^n$ and using the
definitions of $G, \Omega$, one can see that $\Case{1}{2 \hbar}G(\Psi,
\Psi') = \sum_n x^n x^{'n} + y^n y^{'n}$ and $\Case{1}{2
\hbar}\Omega(\Psi, \Psi') = \sum_n x^n y^{'n} - y^n x^{'n}$. Viewing
$(x^n, y^n)$ as (global) `coordinates' on $\cal{H}$ one can see that $G$
corresponds to the infinite order identity matrix while $\Omega$
corresponds to the infinite dimensional analogue of the canonical form
of a symplectic matrix. The coordinates are Darboux coordinates with
$y^n, x^n$ as generalized coordinates and momenta, respectively. A
change of basis effected by a unitary transformation just corresponds to
an orthosymplectic transformation, analogue of the result that $U(n,
\mathbb{C})$ is isomorphic to $OSp(2n, \mathbb{R})$. A separable Hilbert
space can thus be viewed as the $N \to \infty$ form of the usual phase
space $\mathbb{R}^{2 N}$. A non-separable Hilbert space does not admit a
countable basis and an identification as above is not possible.  Much of
the finite dimensional intuition from $\mathbb{R}^{2 N}$ can be borrowed
for separable Hilbert spaces.

The real vector space can naturally be thought of as a (infinite
dimensional) manifold with tangent spaces at each point being identified
with the vector space itself. Explicitly, a tangent vector
$|\Phi\rangle$ at a point $\Psi \in \cal{H}$ acts on real valued
functions, $f(\Psi): \cal{H} \to \mathbb{R}$ as, 
\begin{equation}
|\Phi\rangle|_{\Psi}(f) ~:=~ \lim_{\epsilon \to 0} \Case{f(\Psi + \epsilon
\Phi) - f(\Psi)}{\epsilon}
\end{equation}

Clearly, every vector in $\cal{H}$ can be viewed as a {\em constant}
vector field on $\cal{H}$ viewed as a manifold. The $\Omega$ introduced
above is then a non-degenerate 2-form on $\cal{H}$ which is trivially
closed and hence defines a symplectic structure on $\cal{H}$.  This
immediately allows one to define, for every once differentiable
function $f(\Psi)$, a corresponding {\em Hamiltonian vector field},
$X_f$, via the equation: $\Omega(X_f, Y) = Y(f) \ \forall $ vector
fields $Y$. Non-degeneracy of $\Omega$ implies that Hamiltonian vector
fields are uniquely determined. The Poisson bracket between two such
functions, $f, g$, is then defined as: $\{f, g\}_q := \Omega(X_f, X_g)$.

Next, for every self adjoint operator $\hat{F}$, we get a non-constant
vector field $X_F|_{\Psi} := -\Case{i}{\hbar} \hat{F}|\Psi\rangle$ and
in analogy with the Schrodinger equation, it is referred to as a
Schrodinger vector field. This vector field turns out to be a {\em
Hamiltonian vector field} i.e. there exist a function $f(\Psi)$ such
that $\Omega(X_F, Y) = df(Y) := Y(f)$ for all vector fields $Y$.  From
the definition of $X_F$ and of $\Omega$ in terms of the inner product,
it follows that $\Omega(X_F, Y)|_{\Psi} = \langle \hat{F} \Psi,
Y|_{\Psi}\rangle + \langle Y|_{\Psi}, \hat{F} \Psi \rangle$.  This is
exactly equal to $Y(f)$ for all vector fields $Y$ {\em provided} we
define $f(\psi) := \langle \Psi, \hat{F} \Psi \rangle$.  Thus, every
self-adjoint operator defines a Schrodinger vector field which is also
Hamiltonian with respect to the symplectic structure and the
corresponding `Hamiltonian function' is the expectation value (up to the
norm of $\Psi$) of the operator. This function is quadratic in its
argument and is invariant under multiplication of $\Psi$ by phases. We
will mostly be concerned with such quadratic functions. For any two such
quadratic functions, $f(\Psi) := \langle \Psi, \hat{F} \Psi\rangle,
g(\Psi) := \langle \Psi, \hat{G} \Psi \rangle$, the Poisson bracket
$\{f, g\}_q = \Omega(X_F, X_G)$ evaluates to $\langle \Case{1}{i
\hbar}\left[\hat{F}, \hat{G}\right] \rangle$. It follows that the
quantum mechanical evolution of expectation value functions is exactly
given by a Hamiltonian evolution:
\begin{equation}
\frac{d}{d t} f(\Psi) = \frac{d}{dt} \langle \Psi, \hat{F} \Psi \rangle
= \langle \Psi, (i \hbar)^{-1}\left[\hat{F}, \hat{H} \right] \Psi
\rangle = \left\{f(\Psi), h(\Psi)\right\}_q
\end{equation}
This shows how the quantum dynamics in a Hilbert space can be viewed as
Hamiltonian dynamics in an infinite dimensional symplectic manifold.

Quantum mechanical state space consists of {\em rays} or elements of the
projective Hilbert space. One can import the Hamiltonian framework to
the projective Hilbert space as well \cite{AshtekarSchilling,Schilling}.
In the first step restrict attention to the subset $\cal{S}$ of
normalized vectors (the quadratic functions defined above now exactly
become the expectation values). The projective Hilbert space can then be
viewed as equivalence classes of the relation: $\Psi \sim \Psi'
\Leftrightarrow \Psi' = e^{i \alpha} \Psi$. There is a natural
projection $\rho: \Psi \in {\cal{S}} \to \left[\Psi\right] \in
{\cal{P}}$ and the natural inclusion $i:\cal{S} \to \cal{H}$. With the
inclusion map we can pull back the symplectic form $\Omega$ to $\cal{S}$
on which it is {\em degenerate}. Furthermore, the degenerate subspace
gets projected to zero under $\rho_*$ and hence, there exist a {\em
non-degenerate} 2-form $\omega$ on $\cal{P}$ such that $i^* \Omega =
\rho^* \omega$. This is also closed and thus endows the projective
Hilbert space with a symplectic structure
\cite{AshtekarSchilling,Schilling}. The quadratic functions defined on
$\cal{H}$ are automatically phase invariant and thus project down
uniquely to functions on $\cal{P}$. In particular, we also get Poisson
brackets among the projected quadratic function satisfying $\{f, g\}_q =
\omega(X_f, X_g)$. In this manner, one obtains a Hamiltonian description
of quantum dynamics.

\subsection{Quotienting Procedure}

Now consider a set of self-adjoint operators $\hat{F}_i, i = 1, \cdots,
n$ with the corresponding quadratic functions $f_i: {\cal P} \to \mathbb{R}$.
Define an equivalence relation on $\cal{P}$ which identifies two states
if they have the same expectation values, $f_i$, of all the $\hat{F}_i$
operators. Let $\Gamma$ denote the set of equivalence classes.  The
values $x^i = f_i(\Psi)$ naturally label the points of $\Gamma$. We will
assume that $\Gamma$ can be viewed as a region in $\mathbb{R}^n$ with
$x^i$ serving as coordinates. In other words, the topology and manifold
structure on $\Gamma$ is assumed to be inherited from the standard ones
on $\mathbb{R}^n$.  The manifold structure of the quantum phase space is
discussed in \cite{Schilling}.  There is a natural projection $\pi:
{\cal P} \to \Gamma$. 

We have two natural subspaces of $T_{\Psi}{\cal P}$: (a) the {\em
vertical} subspace ${\cal V}_{\Psi}$ consisting of vectors which project
to zero i.e. $\pi_* v_{\Psi} = 0 \in T_{\pi(\Psi)} \Gamma$ and (b) the
subspace $({\cal{V}}_F)_{\Psi}$ which is the span of the Hamiltonian
vector fields $\{X_{f_i}, i = 1, \cdots, n\}$.  For notational
simplicity, we will suppress the suffix $\Psi$ on these subspaces. The
vertical vectors naturally annihilate functions which are constant over
the fibre and are thus tangential to the fibre.

Let ${\cal V}_F^{\perp}$, denote the $\omega$-complement of ${\cal
V}_F$ i.e. $v \in {\cal V}_F^{\perp}$ implies that $\omega(X_{f_i},
v) = v(f_i) = 0 \ \forall \ i$. Since $f_i$ are constant over a
fibre, every vertical vector satisfies this condition and thus
belongs to ${\cal V}_F^{\perp}$. Are there vectors in ${\cal
V}^{\perp}$ which are {\em not} vertical? This is a little subtle. It
is easy to see that a vector in ${\cal V}_F^{\perp}$ annihilates all
functions polynomial in $f_i$.  One can consider a function algebra
generated by $f_i$ with suitable restrictions on $f_i$ (and hence on
$\hat{F}_i$) and completed with some suitable norm.  Then all
elements of such a function algebra will also be annihilated by
elements of ${\cal V}_F^{\perp}$. All these functions are of course
constant over the fibre. If the class of functions annihilated by the
vertical vectors coincides with the function algebra, then the
subspace ${\cal V}_F^{\perp}$ coincides with the vertical subspace
otherwise ${\cal V} \subset {\cal V}_F^{\perp}$. We will ignore such
fine prints and simply assume that appropriate choices can be made so
that the vertical subspace {\em is} the $\omega$-complement of ${\cal
V}_F: {\cal V} = {\cal V}_F^{\perp}$. Using this identification, we
will refer to ${\cal V}^{\perp}_F$ as the vertical subspace. At this
stage, we do {\em not} know if ${\cal V}_F$ is symplectic subspace or
not. If and only if ${\cal V}_F$ is symplectic (i.e. $\omega$
restricted to ${\cal V}_F$ is non-degenerate), one has (i) ${\cal
V}_F \cap {\cal V}_F^{\perp} = \{0\}$ {\em and} (ii) $T_{\Psi}({\cal
P}) = {\cal V}_F \oplus {\cal V}_F^{\perp}$.  If, however, ${\cal
V}_F$ is {\em not} symplectic, then the intersection of the
symplectic complements is a non-trivial subspace {\em and} the {\em
vector sum} of the two is a {\em proper} subspace of the tangent
space\footnote{This follows by noting that for any non-trivial
subspace, $W$, of a symplectic space, $V$, $(W + W^{\perp})^{\perp} =
W \cap W^{\perp}$ and $V^{\perp} = \{0\}$.}. In the symplectic case,
it is appropriate to refer to ${\cal V}_F$ as the {\em horizontal}
subspace.  In the non-symplectic case, this terminology for ${\cal
V}_F$ is in-appropriate since there are tangent vectors which are not
in ${\cal V}_F + {\cal V}_F^{\perp}$, are also intuitively
`horizontal' or transversal to the fibre.  In general, we will refer
to vector fields valued in ${\cal V}_F$ as {\em basal vector fields}.

These two cases are precisely distinguished by the non-singularity and
singularity of the matrix of Poisson brackets, $A_{ij} :=
\omega(X_{f_i}, X_{f_j}) = \{f_i,f_j\}_q$ respectively.  This is
because, if the matrix $A_{ij}$ is singular, then there exist linear
combinations, $Y_a := \alpha^i_a X_{f_i}, a = 1, \cdots, m$, of vectors
of ${\cal{V}}_F$ which are in ${\cal V}_F^{\perp}$ (i.e.  $\alpha^i_a
A_{ij} = 0$ has non-trivial solutions) and thus have a non-trivial
intersection. Clearly, ${\cal V}_F$ is not a symplectic subspace in this
case. 

Suppose now that in addition, $\hat{F}_i$ are closed under the
commutator or equivalently $f_i$ are closed under $\{\cdot, \cdot\}_q$.
Then, independent of the (non-)singularity of $A_{ij}$, the following
statements are true by virtue of the Poisson bracket closure property of
$f_i$ : (i) commutator of two vertical vector fields is a vertical
vector field and hence the vertical subspaces define an integrable
distribution on ${\cal P}$ (this of course does not need the closure
property); (ii) the commutator of two basal vector fields is a basal
vector field because of the Poisson bracket closure and hence the ${\cal
V}_F$ subspaces also define an integrable distribution on ${\cal P}$;
(iii) finally, the commutator of a vertical vector field and a basal
vector field is a vertical vector field since $\omega(X_{f_i}, [v,
X_{f_j}]) = [v, X_{f_j}](f_i) = v(\{f_j, f_i\}_q) = 0$. Thus,
${\cal{L}}_v X \in {\cal V}_F^{\perp} , \ \forall X \in {\cal{V}}_F\ ,
\forall v \in {\cal V}_F^{\perp}$. The last equality follows from the
closure property of the Poisson brackets. Due to this last property, the
push-forward $\pi_*$ of $X \in {\cal{V}}_F$ from any point along a
fibre, gives the same (i.e. a well defined) vector field on $\Gamma$.
The vectors $Y_a$ project to the `zero' vector field on $\Gamma$. Thus
$\pi_*({\cal V}_F)$ is a subspace of dimension $(n - m)$ of the tangent
space of $\Gamma$. 

One can make the projection $\pi_*$ explicit in the following manner.
Consider projection of $X_{f_i}$. By definition, for functions $f(x^i)$
on $\Gamma$, $[\pi_* X_{f_i}](f) = X_{f_i}(\pi^* f)$. The pull-back
function is constant over the fibre and it depends on $\Psi$ only
through $f_i(\Psi)$. Hence, 
\begin{equation}
X_{f_i}(\pi^* f) ~=~ \Case{\partial \pi^*(f)}{\partial f_j} X_{f_i}(f_j)
~=~ \{f_i, f_j\}_q \Case{\partial f}{\partial x^j} ~=~
A_{ij}\Case{\partial}{\partial x^j}f.
\end{equation}
Thus, $\xi_i := [\pi_* X_{f_i}] = A_{ij}\Case{\partial}{\partial x^j}$.
It follows immediately that $[\pi_* Y_a] = \alpha^i_a A_{ij}
\Case{\partial}{\partial x^j} = 0$ as noted above. This also implies
that only $(n - m)$ of these vectors in $T_{\vec{x}}(\Gamma)$ are
independent. Noting that $A_{ij}$ being linear combinations of the
$f_i$'s, are constant along the fibres and thus descend to $\Gamma$ as
the same linear combinations of $x^i$. Computing the commutator of the
$\xi_i$ directly and using the Jacobi identity satisfied by the
structure constants, one can verify that the projected vectors fields,
$\pi_* X_{f_i}$ are also closed under commutators (with the same
structure constants). Thus, $\pi_*({\cal V}_F)$ define an integrable
distribution on $\Gamma$. The integral sub-manifolds are candidate {\em
classical phases spaces}, $\Gamma_{cl}$, with a symplectic form $\alpha$
(defined below) satisfying, $\omega = \pi^*\alpha$.

Let $Z_I := \beta_I^i X_{f_i}, \ I = 1, \cdots, (n - m)$ be independent
linear combinations of the vector fields $X_{f_i}$ which are closed
under the commutator bracket and $A_{ij}$ is nonsingular on their span.
Their projections are given as $\zeta_I := [\pi_* Z_I] = \beta_I^i \xi_i
$ and are tangential to the integral sub-manifolds.  If, in addition,
the $\zeta_I$'s commute, then the parameters of their integral curves
provide local coordinates on these integral sub-manifolds. In some
cases (see the examples discussed in the subsections), the integral
sub-manifolds of $\Gamma$ are defined by $m$ equations, $\phi_a(x^i) =
\mathrm{constant}$, with the functions satisfying $\zeta_I(\phi_a) = 0 \
\forall I $. The integral sub-manifolds then are embedded sub-manifolds.
In general, however, one only gets immersed sub-manifold.

Now we would like to define a symplectic form on $\Gamma_{cl}$. This can
be done in two steps. Let $s: \Gamma \to {\cal P}$ be a section. From
this one gets the pull-back, $\tilde{\omega} := s^* \omega$ on $\Gamma$.
This is a closed two form but degenerate if $A_{ij}$ is degenerate.
Since any $\Gamma_{cl}$ is a sub-manifold, we also get a closed two form
$\alpha$ on $\Gamma_{cl}$ via the pull-back of the inclusion map,
$\alpha := i^* \tilde{\omega} = i^* \circ s^* \omega$. Explicitly,
\begin{equation} 
\alpha(\zeta_I, \zeta_J) = \tilde{\omega}(i_*\zeta_I, i_*\zeta_J) =
\tilde{\omega}(\zeta_I, \zeta_J) = \omega(s_*\zeta_I, s_*\zeta_J)\ .  
\end{equation}
Since $\pi_*s_*\zeta_I = \zeta_I$, one sees that $s_*\zeta_I = Z_I +
v_I$ for some $v_I \in {\cal V}_F \cap {\cal V}_F^{\perp}$. Clearly,
$\omega(Z_I + v_I, Z_J + v_J) = \omega(Z_I, Z_J)$ and $\alpha$ is well
defined. That $\omega(v_I, v_J) = 0$ follows because both vectors are
vertical as well as basal.

The 2-form $\alpha$ also turns out to be independent of the section
chosen. To see this, observe that Lie derivative of $\omega$ along a
{\em vertical} vector field $v$, when evaluated on basal vector fields
$X, Y$, vanishes: 
\begin{equation}
[{\cal L}_v \omega](X, Y) = [ d ( i_v \omega ) ](X, Y) = X(\omega(v, Y))
- Y( \omega(v, X)) - \omega(v, {\cal L}_X Y) = 0 \ .
%
\end{equation}
Here we used the facts that the vertical and the basal spaces are
symplectic complements and that the commutator of basal vectors is
basal. Thus, the 2-form $\alpha$ is well defined, independent of
section, closed since it is a pull-back of a closed form and
non-degenerate because $\omega$ is non-degenerate on the subspace
spanned by $Z_I$'s. The definition of $\alpha$ is extended to all vector
fields on the integral sub-manifolds by linearity.  Thus each of the
sub-manifolds is now equipped with a symplectic structure and is a
candidate classical phase space which will be generically denoted as
$\Gamma_{cl}$. Note that the symplectic structure on the integral
sub-manifolds is {\em independent} of the section chosen in the
intermediate steps. It however, depends on the particular $\Gamma_{cl}$,
via the inclusion map.

If $A_{ij}$ is non-singular, $\Gamma$ itself is the classical phase
space. Consider the usual case of $\{\hat{F}_i\} = \{\hat{Q}^a,
\hat{P}_a, \hat{\mathbb{I}}, \ a = 1, \cdots, m\}$ forming the
Heisenberg Lie algebra. The corresponding quadratic functions are
$q^a(\Psi), p_a(\Psi)$ and the constant function with value $1$. The
number of corresponding Hamiltonian vector fields are however one less,
since $X_{^\backprime 1'} = 0$.  Furthermore, the space $\Gamma$ is {\em
a single} hyper-plane in $\mathbb{R}^{2m + 1}$. We can thus take $\Gamma
= \mathbb{R}^{2m}$ and focus only on the non-trivial functions. The
matrix of Poisson brackets is then non-singular. There are no vectors
which are both basal and vertical and $\Gamma$ itself is the classical
phase space.  This case is discussed in detail in
\cite{AshtekarSchilling,Schilling}.  

In summary: In this subsection, we have seen that for every Lie algebra
defined by the self adjoint operators $\hat{F}_i$, one can develop a
natural quotienting procedure and construct a classical phase space
$\Gamma_{cl}$.  The classical phase space is in general an immersed
sub-manifold of the space of equivalence classes, $\Gamma$.  This
procedure is capable of yielding linear as well as non-linear classical
phase spaces.  The linear versus non-linear cases are distinguished by
the (non-)singularity of the matrix of Poisson brackets, $A_{ij}$.  The
vector fields on $\Gamma$ which connect different $\Gamma_{cl}$ are
projections of vector fields which are valued in $T_{\Psi}({\cal P}) -
({\cal V}_F + {\cal V}_F^{\perp})$.  The choice of the basic operators
is naturally made if the Hilbert space itself is obtained as a
representation of the corresponding Lie group (as one would do in the
reverse process of quantization). If however, the Hilbert space is not
so chosen, then the choice is to be made appealing to the purpose of
constructing a classical phase space. Our reason for constructing a
classical phase space has been to develop a semiclassical approximation
and this involves dynamics as well. Thus, the choice of algebra will be
dictated by the quantum dynamics and its semiclassical approximation
sought.

The general procedure given above is illustrated in three simple
examples in the next three subsections.

\subsection{Spin-j system}
The Hilbert space is complex $2j + 1$ dimensional or real (4j + 2)
dimensional and the projective Hilbert space is $\mathrm{CP}^{2j}$. As
basic operators we choose three hermitian matrices $S_i$ satisfying
$[S_i, S_j] = i \hbar \epsilon_{ijk}S_k$.  We also have the relation,
$\langle\Psi|\sum \hat{S}_i^2 |\Psi\rangle = j(j + 1)\hbar^2$.
Furthermore, $\langle \hat{S}_i^2\rangle = (x^i)^2 + \Delta S_i^2 \ge
(x^i)^2$.  Therefore, $r^2 := \sum (x^i)^2 \le j(j + 1)\hbar^2$. 

The equivalence relation is defined in terms of $S^i(\Psi) :=
\Psi^{\dagger} S_i \Psi$ which gives a 3 dimensional $\Gamma$ and ${\cal
V}_S$ subspaces of tangent spaces of the $CP^{2j}$. Correspondingly, the
symplectic structure gives $\omega(X_{S_i}, X_{S_j}) = \{S^i, S^j\}_q =
\epsilon_{ijk}S^k$.  Clearly ${\cal V}_S$ cannot be a symplectic
subspace and indeed $Y := S^i(\Psi)X_{S_i}$ is also a vector field
valued in ${\cal V}_S^{\perp}$. Denoting the coordinates on $\Gamma$ by
$x^i = S^i$, the projections are given by, $\xi_i := \pi_* X_{S_i} =
\epsilon_{ijk}x^k\partial_j$. These projected vectors are {\em not}
independent.  Let $\vec{\beta}_I, \ I = 1, 2$ be any two, 3 dimensional
vectors and define $\zeta_I := \beta_I^i \xi_i$. Their commutator is
given by $[ \vec{x}\cdot\vec{\beta}_J \ \beta_I^i\partial_i -
\vec{x}\cdot\vec{\beta}_I \ \beta_I^i\partial_i ]$. Clearly, if the two
vectors $\vec{\beta}_I$ are orthogonal to the `radial vector' $\vec{x}$,
then the two vector fields commute. Let us further choose the two
vectors $\vec{\beta}_I$ to be mutually orthogonal in anticipation.  The
two dimensional integral sub-manifolds are defined by $\phi(\vec{x}) =
\phi(\sum x^ix^i) = \mathrm{constant}$, which are 2-spheres as expected
for $\Gamma_{cl}$. Note that this is true for all spins. 

Let us compute the induced symplectic structure. The 2-form $\alpha$ is
defined by,
\begin{equation}
\alpha(\zeta_I, \zeta_J) := \beta_I^i \beta_J^j \tilde{\omega}(\xi_i,
\xi_j) = \beta_I^i \beta_J^j \omega(X_{S_i}, X_{S_j}) = \beta_I^i
\beta_J^j \epsilon_{ijk} S^k = \vec{r}\cdot \vec{\beta}_I \times
\vec{\beta}_J.
\end{equation} 

Thus we have several spheres of radii $r$ as candidate classical phases
spaces and the normalization of the induced symplectic structure depends
on both the particular sphere as well as arbitrary magnitudes of the two
$\vec{\beta}_I$. The arbitrariness due to the magnitudes can be fixed by
the choice of coordinates provided by the integral curves of the
commuting vector fields.  The normalization of $\beta_I$ can be deduced
by requiring $\zeta_1 := \partial_{\theta}, \ \zeta_2 :=
\partial_{\phi}$ where $\theta, \phi$ are the usual spherical polar
tangles. This requirement together with orthogonality of $\vec{r},
\vec{\beta}_I$, fixes the normalizations completely and leads to
$\alpha(\partial_{\theta}, \partial_{\phi}) = r \ \mathrm{sin}
(\theta)$. 

For the case of $j = 1/2$, the Hilbert space is four (real) dimensional,
the subset of normalized vectors, ${\cal S}$ is the three dimensional
sphere and the state space ${\cal P}$ is the two dimensional sphere. The
space $\Gamma$ itself becomes {\em two} dimensional, the fibre over each
point of $\Gamma$ is zero dimensional. Thus there are no vertical
vectors and the subspace ${\cal V}_S$ is {\em two} dimensional and
symplectic. Thus $\Gamma_{cl} = \Gamma$ holds. 

\subsection{Particle on a circle} In this case the Hilbert space is the
space of periodic (say) square integrable complex functions on the
circle. We have three natural operators forming a closed commutator
sub-algebra:
\begin{eqnarray}
\langle \phi | n \rangle & := & \frac{1}{\sqrt{2 \pi}} e^{i n \phi},
\hspace{1.0cm} n \in \mathbb{Z} \hspace{2.0cm} \mathrm{basis ~
functions} \\
\widehat{\mathrm{cos}}| n\rangle & := & \frac{1}{2}\left\{ |n + 1
\rangle + |n - 1 \rangle \right\}; \label{BasicDefinitions}\\
\widehat{\mathrm{sin}}| n\rangle & := & \frac{1}{2i}\left\{ |n + 1
\rangle - |n - 1 \rangle \right\}; \nonumber \\
\widehat{\mathrm{p}}| n\rangle & := & \hbar n |n \rangle \nonumber \\
\left[ \widehat{\mathrm{cos}}, ~ \widehat{\mathrm{sin}}\right] & = & 0
\label{BasicCommutators}\\
\left[ \widehat{\mathrm{cos}}, ~ \widehat{\mathrm{p}}\right] & = & - i
\hbar ~ \widehat{\mathrm{sin}} \nonumber \\
\left[ \widehat{\mathrm{sin}}, ~ \widehat{\mathrm{p}}\right] & = & i
\hbar ~ \widehat{\mathrm{cos}} \nonumber \\
\widehat{\mathrm{cos}} \ \widehat{\mathrm{cos}}  +
\widehat{\mathrm{sin}}\ \widehat{\mathrm{sin}} & = & \mathbb{I}
\label{TrigId}
\end{eqnarray}
Thus, for $\hat{F_i}$ we choose the $\widehat{\mathrm{cos}}, ~
\widehat{\mathrm{sin}}$ and $\hat{p}$ operators and denote the
corresponding quadratic functions $f_i(\Psi)$ by cos($\Psi$),
sin($\Psi$) and p($\Psi$).  The Poisson brackets among these is obtained
via $ \{f_i, f_j\}_q = \omega(X_{f_i}, X_{f_j}) = \langle
\Case{1}{i\hbar}[\hat{F}_i, \hat{F}_j ] \rangle$. From the
(\ref{TrigId}) it is easy to see that 
\begin{equation} \label{Bounds}
\langle\mathrm{cos}^2\rangle + \langle\mathrm{sin}^2\rangle  ~=~ \langle
\mathrm{cos} \rangle^2 + \langle \mathrm{cos} \rangle^2 + (\Delta
\mathrm{cos})^2 + (\Delta\mathrm{sin})^2 ~ = ~ 1.
\end{equation}

The space $\Gamma$ defined as the set of equivalence classes of rays
having the same expectation values of the three basic operators, is the
region in $\mathbb{R}^3$ : $(-1 \ < \ x \ < \ 1, -1 \ < \ y \ < \ 1,
-\infty \ < \ z \ < \ \infty)$ where $x = \mathrm{cos}, y =
\mathrm{sin}, z = \mathrm{p}$ and $0 \le x^2 + y^2 < 1$. Notice that the
uncertainties in the trigonometric operators are also bounded from
above.

It is easy to see that the span of the three Hamiltonian vector fields
on the projective Hilbert space, $X_{cos}, X_{sin}, X_p$ also contains a
vector field, $Y = \mathrm{cos} X_{cos} + \mathrm{sin} X_{sin}$, which
is valued in ${\cal V}_F^{\perp}$. From $\pi_* X_{f_i} = \{f_i, f_j\}_q
\Case{\partial}{\partial x^j}$ it follows that the projection of these
vectors fields to $\Gamma$ are given by,
\begin{equation}
\pi_* X_{cos} ~ = ~ - y \frac{\partial}{\partial z}~~,~~ \pi_* X_{sin} ~
= ~  x \frac{\partial}{\partial z}~~,~~ \pi_* X_{p} ~ = ~  -x
\frac{\partial}{\partial y} ~+~ y \frac{\partial}{\partial x}~~,~~ \pi_*
Y ~=~ 0\ .
\end{equation}
Clearly all three projected vectors at any point of $\Gamma$ are not
independent and the `radial' vector field $x\partial_x + y \partial_y$
on $\Gamma$ is not a projection of any vector field valued in ${\cal
V}_F + {\cal V}_F^{\perp}$. As an example, the two independent vectors
$\zeta_1 := \pi_*(- \mathrm{sin} X_{cos} + \mathrm{cos} X_{sin})$ and
$\zeta_2 := \pi_* X_p$ commute. The two dimensional vector spaces
spanned by these, define a two dimensional integrable distributions on
$\Gamma$.  Introducing the usual polar coordinates in the $x-y$ plane,
$x := r \mathrm{cos} \phi, y := r \mathrm{sin} \phi$, the two vector
fields can be expressed as $\zeta_1 = r^2 \partial_z, \ \zeta_2 = -
\partial_{\phi}$. The integral sub-manifolds are defined by the integral
curves of these vector fields and are characterized by $r =
\mathrm{constant} (\neq 0)$. Thus the classical phase space manifolds
are cylinders as expected.

To obtain a symplectic structure $\alpha$, on any of these cylinders,
define $\alpha(\zeta_1, \zeta_2) := \tilde{\omega}(- \mathrm{sin}
X_{cos} + \mathrm{cos} X_{sin}, X_p) =  - \mathrm{sin} \{\mathrm{cos},
\mathrm{p}\} + \mathrm{cos} \{\mathrm{sin}, \mathrm{p}\} =
\mathrm{sin}^2 + \mathrm{cos}^2  = r^2 $. This is very much like the
usual symplectic structure on a cylinder. Indeed, the polar coordinates
introduced suggest that $\phi$ and $p_{\phi} := z/r^2$ be identified as
the usual canonical variables on the cylinder. 

For $r = 0$, of course one does not have a classical phase space (the
cylinder degenerates to a line). Since we are interested in constructing
classical phases spaces with a view to constructing a semiclassical
approximation, we exclude the degenerate case. This also means that we
exclude the fibres over the line $(x = 0 = y)$. These fibres consists of
all normalized states of the form $|\Psi\rangle = \sum_{n \in
\mathbb{Z}} \psi_n|n\rangle$ with the coefficients satisfying the
condition that if $\Psi_n \neq 0$ for some $n$ then $\Psi_{n \pm 1} =
0$. Analogous considerations will be more relevant in the next example.

In both the examples above, the Hilbert space has been separable and the
classical phase spaces are obtained as embedded sub-manifolds of
$\Gamma$. Both these features will change in the next example.

\subsection{Isotropic loop quantum cosmology}

This is an example with a non-separable Hilbert space \cite{ABL,Shadow}
but in many ways still similar to the particle on a circle example. We
can use the so-called triad representation whose basis states are
labelled by eigenvalues of the self-adjoint triad operator $\widehat{p}$
which take {\em all} real values with corresponding eigenstates being
normalized.  Unlike the particle on a circle example, the Hilbert space
is not made up of {\em periodic} (or quasi-periodic) functions on a
circle, but consists of {\em almost periodic} functions of a real
variable $c$. The Hilbert space carries a {\em non-weakly-continuous}
unitary representation of the Heisenberg group such that there is no
operator $\widehat{c}$ generating translations of eigenvalues of
$\widehat{p}$.  However, exponentials of $i c$ are well defined
operators. In this regard, this is similar to the particle on a circle.
Thus we can define:
\begin{eqnarray}
\langle c | \mu \rangle & := & \frac{1}{\sqrt{2 \pi}} e^{i \mu c},
\hspace{1.0cm} \mu \in \mathbb{R} \hspace{2.0cm} \mathrm{basis ~
functions} \\
\widehat{\mathrm{cos}}_{\lambda}| \mu\rangle & := & \frac{1}{2}\left\{
|\mu + \lambda \rangle + |\mu - \lambda \rangle \right\};
\label{BasicDefinitionsTwo}\\
\widehat{\mathrm{sin}}_{\lambda}| \mu\rangle & := & \frac{1}{2i}\left\{
|\mu + \lambda \rangle - |\mu - \lambda \rangle \right\}; \nonumber \\
\widehat{\mathrm{p}}| \mu\rangle & := & \hbar \mu |\mu \rangle \nonumber
\\
\left[ \widehat{\mathrm{cos}}_{\lambda}, ~
\widehat{\mathrm{sin}}_{\lambda}\right] & = & 0
\label{BasicCommutatorsTwo}\\
\left[ \widehat{\mathrm{cos}}_{\lambda}, ~ \widehat{\mathrm{p}}\right] &
= & - i \hbar \ \lambda\ ~ \widehat{\mathrm{sin}}_{\lambda} \nonumber \\
\left[ \widehat{\mathrm{sin}}_{\lambda}, ~ \widehat{\mathrm{p}}\right] &
= & i \hbar \ \lambda\ ~ \widehat{\mathrm{cos}}_{\lambda} \nonumber \\
\widehat{\mathrm{cos}}_{\lambda} \ \widehat{\mathrm{cos}}_{\lambda}  +
\widehat{\mathrm{sin}}_{\lambda}\ \widehat{\mathrm{sin}}_{\lambda} & = &
\mathbb{I} \label{TrigIdTwo}
\end{eqnarray}

The label $\lambda$ indicates that we can define such `trigonometric'
operators for every {\em non-zero real} number $\lambda$. Since
$\widehat{\mathrm{cos}}_{-\lambda} = \widehat{\mathrm{cos}}_{\lambda}$
and $\widehat{\mathrm{sin}}_{-\lambda} =
-\widehat{\mathrm{sin}}_{\lambda}$, we will take $\lambda > 0$. We will
denote by ${\cal L}_{\lambda}$, the Lie algebra defined by
(\ref{BasicCommutatorsTwo}) for any fixed $\lambda$. In the circle
example also we could define such operators, but the periodicity would
restrict the $\lambda$ to integers.  The trigonometric operators for
different $\lambda$'s of course commute.  The Hilbert space carries a
{\em reducible} representation of the Lie algebra ${\cal L}_{\lambda}$
for every $\lambda$ while every (countable) subspace ${\cal H}_{a,
\lambda}$, spanned by vectors of the form $\{|a + k \lambda\rangle, k
\in \mathbb{Z}\}$, gives an irreducible representation of ${\cal
L}_{\lambda}$, for every $a \in [0, \lambda)$.

The presence of $\lambda$ leads to $\zeta_1 = \lambda r^2 \partial_z,
\zeta_2 = - \lambda \partial_{\phi}$. The integral curves are defined by
constant $r$ and $\dot{\phi} = -\lambda$. The integral curves of
$\zeta_2$ are clearly closed and we can choose the scale of $\zeta_2$ so
that the curve parameter itself is $\phi$, i.e. redefine $\zeta_2 := -
\lambda^{-1} \zeta_2$.  We see that the classical phase space is the
cylinder as in the previous example.

But the expected classical phase space is supposed to be $\mathbb{R}^2$
whose Bohr quantization leads to the non-separable quantum Hilbert
space.  How does one see the linear classical phase space? 

Now, unlike the previous example where the trigonometric operators must
induce shifts by integers to respect the periodicity, here we have more
possibilities.  We could consider enlarging the set of basic operators
by including $\widehat{\mathrm{cos}_{\lambda}},
\widehat{\mathrm{sin}_{\lambda}}$ for $n$ distinct values of $\lambda$.
Thus our space of equivalence classes, $\Gamma$ will be $(2 n + 1)$
dimensional. One can see easily that there are now $(2 n - 1)$
independent vectors in ${\cal V}_F$ which will project to zero vector on
$\Gamma$ and we will be left with exactly {\em two} independent vector
fields, closed under commutator, on $\Gamma$.  Explicitly, these vector
fields can be chosen as: $\zeta_1 = \pi_*( \sum_i \lambda_i\{ -
\mathrm{sin}_{\lambda_{i}}X_{cos_{\lambda_{i}}} +
\mathrm{cos}_{\lambda_{i}}X_{sin_{\lambda_{i}}} \} )\ = \sum_i \lambda_i
\{ ( x^i)^2 + (y^i)^2 \} \partial_z$ and $\zeta_2 = \pi_* X_p = \sum_i
\{ - \lambda_{i} ( x^i \partial_{y^i} - y^i\partial_{x^i} ) \}$. There
is freedom available in defining $\zeta_1$, but $\zeta_2$ has freedom of
{\em only} overall scaling. In the $(x^i, y^i)$ planes, we can
introduce the polar variables $(r_i, \phi_i)$ and see that the integral
curves of $\zeta_1$ are along the $z$-direction while those of $\zeta_2$
are defined by the equations: $r_i = \mathrm{constant}_i (\neq 0),
\dot{\phi^i} = - \lambda_i$ i.e. winding curves on an $n$-torus, $T^n$.
Evidently, for $\lambda_i$ with irrational ratios, these integral curves
are non-periodic.  Thus we get a map of $\mathbb{R}$ into the $T^n$
which however cannot be an embedding since the induced topology on the
winding curve is not the standard topology on $\mathbb{R}$. One has only
an immersion. Combining with the curve parameter of $\zeta_1$, one has
an immersion of $\mathbb{R}^2$ and this is adequate to define the
symplectic form $\alpha$. In effect we obtain the classical phase space
which is topologically $\mathbb{R}^2$ and is immersed in an
$\mathbb{R}\times T^n \subset \Gamma$. 

The symplectic form is computed as before and leads to, $\alpha(\zeta_1,
\zeta_2) = \sum_i \lambda^2_i r_i^2$. Denoting the curve parameter for
the vector field $\zeta_2$ by $Q$ and defining its canonically conjugate
variable as $P := z/(\sum_i \lambda^2_i r_i^2)$, one has the usual
symplectic form on $\mathbb{R}^2$.

If {\em all} the $\lambda_i$ happen to be rational numbers (or rational
multiples of a single irrational number), then the integral curves of
$\zeta_2$ are closed and one would obtain the classical phase space to
be the cylinder (or a covering space thereof). Clearly, to obtain the planar phase space, one must
have at least {\em two} $\lambda$' whose ratio is an irrational number
{\em with the corresponding $r_i \neq 0$}. 

Here we find an example where depending upon the choice of basic
operators or more precisely the parameter(s) $\lambda_i$, we can obtain
{\em two different topologies} for the classical phase space.  We also
see that although the choice of basic operators can vary widely, one
always obtains a {\em two dimensional} classical phase space which is
generically $\mathbb{R}^2$.

All these mathematical procedures of constructing `classical' phase
spaces from the quantum one become physically relevant for studying
semiclassical approximation(s) only when {\em a dynamics} is stipulated
and attention is paid to the uncertaintities in the basic operators
$\hat{F}_i$. These ultimately decide which choice of the Lie algebra of
$\hat{F}_i$ is useful. We address these in the next section.

\section{Dynamics and semiclassicality}
In this section, we discuss how a quantum evolution taking
place in the quantum phase space can generate a Hamiltonian
evolution on a classical phase space constructed in the
previous section. This will naturally involve a choice of a
`section'. While, every section will generate {\em a}
classical dynamics, further conditions have to be imposed for
the classical dynamics to be a good approximation to the
quantum dynamics.  These are the conditions of
`semiclassicality' and `tangentiality'. The fact that the
classical phase of the isotropic cosmology arises as an
immersed manifold in the base space, introduces further
considerations which are discussed next.

So far we just constructed a `classical' phase space, $\Gamma_{cl}$,
selecting a sub-algebra of self-adjoint operators and using the
naturally available symplectic geometry of the quantum state space. This
is completely independent of any dynamics. Consider now a quantum
dynamics specified by a self-adjoint Hamiltonian operator, $\hat{H}$.
The bundle structure allows one to project any quantum trajectory
generated by $X_{H}$, to a trajectory in the base space $\Gamma$. 

In the general case where a $\Gamma_{cl}$ is a submanifold  of $\Gamma$,
the first problem is that projected trajectories may not even be
confined to $\Gamma_{cl}$ i.e.  $\pi_*(X_H) \notin \pi_*({\cal V}_F)$.
Notice that $X_H$ is not `constant' along the fibres (${\cal L}_v X_H$
is not vertical) and therefore we need to specify the points on the
fibres.  If there are {\em no} quantum trajectories whose projections
remain confined to some $\Gamma_{cl}$, then the choice of the Lie
algebra of $\hat{F}_i$ used in the quotienting procedure, {\em in
conjunction with the Hamiltonian}, is inappropriate for developing a
semiclassical approximation and one has to look for a different choice.
Let us assume that one has gone through this stage and found a suitable
Lie algebra so that there are at least some quantum trajectories whose
projections are confined to $\Gamma_{cl}$. Note that when $\Gamma_{cl} =
\Gamma$, as in the usual case of Heisenberg Lie algebra, this issue does
not arise at all.

The second issue is whether there are `sufficiently many' quantum
trajectories whose projections are confined to $\Gamma_{cl}$.
Sufficiently many would mean that projection of these trajectories will
be at least an open set of $\Gamma_{cl}$. This in turn allows the
possibility that at least a portion of the classical phase space can be
used for a semiclassical approximation. Let us also assume this.

The third issue is whether the projected trajectories are generated by a
Hamiltonian function on $\Gamma_{cl}$ and whether such a Hamiltonian
function can be constructed from the quantum Hamiltonian function
$H(\Psi) = \langle \Psi|\hat{H}|\Psi\rangle$ already available on the
quantum state space.  This function however is not constant along the
fibres and cannot be `projected' down to the base space. One has to
choose a section: $s:\Gamma \to {\cal P}, ~ \pi \circ s = id$. There are
infinitely many choices possible.  However, for every such choice, we
can define a function $\tilde{H}: \Gamma \to \mathbb{R}$ as the pull
back of $\langle\Psi|\hat{H}|\Psi\rangle|_{s(\Gamma)}$. The restrictions
of these functions to any of the sub-manifolds, $\Gamma_{cl}$ would be
candidate {\em effective Hamiltonians}, $H_{\mathrm{eff}}$.

In this regard, we would like to note a simple fact. Let $s: \Gamma \to
{\cal P}$ be a section (possibly local) and let $X_H$ be a Hamiltonian
vector field on ${\cal P}, H(\Psi) := \langle\Psi| \hat{H}
|\Psi\rangle$. Let $\xi := \pi_* (X_H|_{s(\Gamma)})\ , ~\tilde{H}(x^i)
:= s^*(H|_{s(\Gamma)})$. The (symplectic) 2-forms are related as $\omega
:= \pi^* \tilde{\omega} \leftrightarrow \tilde{\omega} = s^* \omega$.
Then it is true that $\xi$ is a Hamiltonian vector field (on $\Gamma$)
with respect to $\tilde{\omega}$, with $\tilde{H}$ as the corresponding
function. This follows as,
\begin{equation}
\tilde{\omega}(\xi, \xi_i) ~=~ \omega(X_H, X_{f_i}) ~=~ X_{f_i}(H) ~=~
\pi_*(X_{f_i})(\tilde{H}) ~=~ \xi_i (\tilde{H}) ~~\forall ~ i.
\end{equation}
Since any $\Gamma_{cl}$ is a sub-manifold of $\Gamma$ we have natural
projection (restriction) and inclusion maps so that we obtain the
restriction of $\xi$ to tangent spaces of $\Gamma_{cl}$ as the
Hamiltonian vector field with $i^* \tilde{H}$ as the corresponding
function. The net result is that for any section and any Hamiltonian
vector field, we can always obtain {\em a} classical Hamiltonian
description. Since there are infinitely many sections, we have
infinitely many classical dynamics induced from a given quantum
dynamics.  Is any of these dynamics a `good' approximation to the
quantum dynamics? For this we have to invoke further conditions on the
sections.

The expectation value functions are naturally observables and our
restriction to expectation values of a subset of the observables
$\hat{F}_i$, amounts to saying that in a given situation and with a
given experimental capability, we can discern the quantum dynamics only
in terms of these few `classical' variables. With better experimental
access, we may need more such variables. Generically, these observables
also have quantum uncertainties and provided these uncertainties are
smaller than the observational precision, we can ignore them, thereby
justifying a classical description.  Obviously, such a property will not
be exhibited by all quantum states and not for all times.  Consequently,
one first defines a quantum state to be semi-classical with respect to a
set of observables $\hat{F}_i$ provided the uncertainties $(\Delta
F_i)_{\Psi}$, are smaller than some prescribed tolerances $\delta_i$.
For such a notions to be observationally relevant/useful, semi-classical
states so defined must evolve, under quantum dynamics, into
semi-classical states for {\em sufficiently long durations}.  
Since we identified a classical phase space using equivalence
relation specified by a set of functions $f_i(\Psi) := \langle\Psi|
\hat{F}_i |\Psi\rangle$, closed under the $\{\cdot, \cdot\}_q$, these
are naturally the candidate observables with respect to which we can
define semiclassicality of a state. 

On each fibre then, we could find subsets for which the uncertainties
would be smaller than some specified tolerances $\delta_i$ \footnote{The
tolerances could be prescribed to vary over different regions of
$\Gamma$. For instance, for larger expectation values, one could have
less precision and thus permit larger values of $\delta_i$.}. Quantum
states within these subsets would be semi-classical states. Let us
assume that we identify such bands of semiclassicality on each fibre.
Our sections should intersect each fibre within its semi-classical
band(s). This still leaves an infinity of choices and is still not
enough to have the classical trajectories generated by an effective
Hamiltonian to approximate the projections of quantum trajectories.  

That the projection of a Hamiltonian vector field restricted to
$s(\Gamma)$ is also a Hamiltonian vector field on $\Gamma_{cl}$ follows
independent of whether or not the Hamiltonian vector field $X_H$ is {\em
tangential} to the section. However, tangentiality of the Hamiltonian
vector field $X_H$ to a section is necessary so that projection of a
quantum trajectory in ${\cal P}$, generated by $X_H$, gives the
corresponding classical trajectory (in $\Gamma$) generated by
$\xi_{\tilde{H}}$. If tangentiality fails, then the pre-image of a
classical trajectory, will {\em not} be a single quantum trajectory i.e.
pre-images of nearby tangent vectors of a classical trajectory will
belong to {\em different} quantum trajectories.  Once $X_H$ is
tangential to $s(\Gamma)$, the classical evolution in $\Gamma$ will
mirror the quantum evolution (and by assumption made at the beginning,
classical evolution will be confined to $\Gamma_{cl}$). 

Thus, to develop a semi-classical approximation what is needed is to
guess or identify a classical phase space $\Gamma_{cl}$ such that
sufficiently many quantum trajectories project into $\Gamma_{cl}$,
select criteria of semiclassicality and {\em choose a suitable (and
possibly local) section which is tangential to the $X_H$ and lies within
a semi-classical band.} The classical phase space is obtained via the
quotienting procedure while the effective classical Hamiltonian is
obtained as the pull-back of $\langle \hat{H} \rangle$. Note that we
have {\em not} assumed that the Hamiltonian operator is an algebraic
function of the basic operators chosen in the quotienting procedure. The
effective Hamiltonian however is always a function on $\Gamma_{cl}$ (and
$\tilde{H}$ is a function on $\Gamma$) by construction. Even if the
Hamiltonian operator is an algebraic function of the basic operator, the
$\tilde{H}$ will {\em not} be an algebraic function of the expectation
values of basic operators. 

Suppose now that we have chosen a section (over $\Gamma$) satisfying all
the conditions above. In the general case where $\Gamma_{cl} \neq
\Gamma$, we have to restrict to the section over a $\Gamma_{cl}$ i.e.
restrict to those states in the section which project into
$\Gamma_{cl}$. Secondly, in general we will not have operators
corresponding to the canonical coordinates in $\Gamma_{cl}$ and hence for
the semiclassical criteria we have to use only the basic operators used
in obtaining $\Gamma$. The LQC example, reveals implications of these
aspects. We will see that the Hamiltonian operator also needs to have
non-trivial dependences on all the basic operators chosen for the
quotienting procedure and the semiclassicality criterion also needs to
be phrased differently.

Recall that to construct the classical phase space $\mathbb{R}^2$ (as
opposed to a cylinder), one needs to use at least two sets of
trigonometric operators with labels $\lambda, \lambda'$ such that (i)
$\lambda'/\lambda$ is irrational and (ii) $r, r'$ both being non-zero.
If a state in a section has support only on a lattice generated by
$\lambda$ (say), or a subset thereof, one can see immediately that
$\mathrm{cos}_{\lambda'} = 0 = \mathrm{sin}_{\lambda'}$ and hence $r' =
0$. Therefore such states in a section will not represent points in the
classical phase space (but will represent points in a cylinder).  Let us
then assume that our section consists of states having support on
(sub-)lattices generated by both $\lambda, \lambda'$. To be definite,
let us take, $|\Psi_{a,\lambda}\rangle$ to be a normalized vector which
is a linear combination of vectors of the form $|a + k\lambda\rangle, k
\in \mathbb{Z}, a \in (0, \lambda)$ and likewise
$|\Psi_{b,\lambda'}\rangle$. Let a state in a section be $|\Psi\rangle
:= (|\Psi_{a,\lambda}\rangle + |\Psi_{b,\lambda'}\rangle)/\sqrt{2}$.  We
are now guaranteed to have the expectation values of $\hat{p},
\widehat{\mathrm{cos}}_{\lambda}, \widehat{\mathrm{sin}}_{\lambda},
\widehat{\mathrm{cos}}_{\lambda'}, \widehat{\mathrm{sin}}_{\lambda'}$ to
determine a unique point of the classical phase space, $\mathbb{R}^2$.
Furthermore, for any basic trigonometric operator with a label
$\lambda''$, with $\lambda''$ irrational multiple of $\lambda,
\lambda'$, its expectation value in the state $|\Psi\rangle$ will be
zero and the uncertainty will be $1/2$. For the uncertainty, we note
the identities, $\widehat{\mathrm{cos}^2}_{\lambda''} = (1 +
\widehat{\mathrm{cos}}_{2\lambda''})/2$ and
$\widehat{\mathrm{sin}^2}_{\lambda''} = (1 -
\widehat{\mathrm{cos}}_{2\lambda''})/2$.  We would like to see evolution
of these quantities generated by a self-adjoint Hamiltonian operator,
$\hat{H}_{\lambda_0}$ which is a function only of $\hat{p},
\widehat{\mathrm{cos}}_{\lambda_0}, \widehat{\mathrm{sin}}_{\lambda_0}$
for some fixed $\lambda_0$.  The quantum evolution of
$\langle\mathrm{cos}_{\lambda''}\rangle,
\langle\mathrm{sin}_{\lambda''}\rangle$ will be given by the expectation
values of the commutators of the corresponding trigonometric operator
with the Hamiltonian operator, 

The Hamiltonian acting on $|\Psi\rangle$ will generate vectors with
labels $k\lambda + l\lambda_0$ and $k\lambda' + l\lambda_0$. The
trigonometric operator acting on $\langle\Psi|$ on the other hand will
generate vectors with labels $k\lambda \pm \lambda''$ and $k\lambda' \pm
\lambda''$. If $\lambda_0, \lambda, \lambda', \lambda''$ are all
incommensurate, then the inner product of these states will be zero and
hence the expectation value of the commutator will be zero.
Consequently, expectation values of these trigonometric operators will
not evolve. By the same logic, their uncertainties also will not evolve.
If the Hamiltonian operator is a sum of a function of $\hat{p}$ alone
plus a function of the trigonometric operator alone (as could happen for
polymer quantization of usual systems \cite{Shadow}), then of course
expectation values and uncertainties of trigonometric operators with
both labels $\lambda, \lambda'$, will evolve, irrespective of
$\lambda_0$. Barring this exception, the only way to get a non-trivial
evolution is to have $\lambda_0$ to equal $\lambda$ or $\lambda'$.
Suppose we choose $\lambda = \lambda_0$ (The quantum Hamiltonian is
given and we can choose the states in the section so as to develop a
semiclassical approximation).  Now, the evolution of
$\langle\widehat{\mathrm{cos}_{\lambda_0}}\rangle$ and
$\langle\widehat{\mathrm{sin}_{\lambda_0}}\rangle$ will be non-trivial
and so also the evolution of the corresponding uncertaintities. However,
the evolution of corresponding quantities for $\lambda'$ will still be
trivial! In effect, the projection of the quantum trajectories on the
classical phase space will {\em not} be non-periodic curves. 

Thus, to have a non-trivial classical evolution (in $\Gamma_{cl}$),
one will have to have (a) trigonometric operators with an
incommensurate set of $\lambda$'s (at least two), (b) a section
satisfying semiclassicality and tangentiality, (c) states (in a
section) involving (sub-)lattices corresponding to these $\lambda$'s
and (d) the quantum Hamiltonian also involving trigonometric
operators with the chosen set of $\lambda$'s.  Note that this is a
statement about developing a semiclassical approximation and does
{\em not} imply or indicate any inconsistency of the choice of the
Hamiltonian operator at the fundamental quantum level.  

Having multiple incommensurate $\lambda$'s also affects the
uncertaintities in the trigonometric operators. Continuing with the
choice of just $\lambda, \lambda'$, one sees that,
\begin{equation}
(\Delta \mathrm{cos}_{\lambda})_{\Psi}^2 = \Case{1}{4} + \Case{1}{4}
\langle\Psi_{a,\lambda}| \widehat{\mathrm{cos}}_{\lambda}
|\Psi_{a,\lambda}\rangle^2 + \Case{1}{2} (\Delta
\mathrm{cos}_{\lambda})_{\Psi_{a,\lambda}}^2 
\end{equation}
The first term comes from $|\Psi_{b,\lambda'}\rangle$ piece of the wave
function.  Similar expressions are obtained for the other three other
trigonometric operators. Notice that these uncertainties are always
larger than $1/4$. If we choose a state based on $N$ incommensurate
$\lambda$'s, then the first term changes to $(N -1)/(2N)$, the
coefficient of the second term changes to $(N -1)/N^2$ while the
coefficient of the third term changes to $1/N$. For large $N$, the
uncertaintities become $1/2$. This is irrespective of the details of the
states. Thus, if the semiclassicality criteria required uncertainties in
the trigonometric operators to be small (recall that uncertainties are
bounded above by 1 (\ref{Bounds})), then {\em no} state involving
several incommensurate $\lambda$'s will satisfy the criterion of
semiclassicality. Since for $N = 1$ (single $\lambda$), only the last
term survives, a way out is to require the uncertainties in
$\widehat{\mathrm{cos}}_{\lambda}, \widehat{\mathrm{sin}}_{\lambda}$
{\em in the states of the form $|\Psi_{a,\lambda}\rangle$} to be smaller
than prescribed tolerances. This extra feature arises because the
classical phase space $\Gamma_{cl}$, is immersed in a complicated way
in the base space $\Gamma$ and one does not have operators corresponding
to the canonical coordinates on $\Gamma_{cl}$ which could be used in
formulating semiclassicality criteria.

\section{Discussion}
In this work, we have explored a particular strategy of developing a
semiclassical approximation, namely, systematically, constructing a
classical Hamiltonian system using the available quantum structures such
that at least some quantum motions can be faithfully viewed as classical
motions within certain precisions. This was done by exploiting the
symplectic structure of the quantum state space which is typically
infinite dimensional. One first constructs a (finite) dimensional
classical phase space by a quotienting procedure which views the quantum
state space as a bundle over a base space $\Gamma$. There are several
such phase spaces one can construct depending on the choice of a Lie
algebra of basic operators.  Generically, one gets classical phase
space, $\Gamma_{cl}$ as a submanifold of the base space. So far,
typically the special case wherein $\Gamma_{cl} = \Gamma$ has been
analysed in the literature, eg
\cite{AshtekarSchilling,Schilling,Shadow,Willis}. We have given a
generalization of the procedure. The main difference that occurs in the
general case is that the vertical subspace and the space spanned by the
Hamiltonian vector fields corresponding to the basic operators, have a
non-trivial intersection and also together they do not span the tangent
space of the bundle. We illustrated the general procedure for three
different types of quantum systems - none resulting from a weakly
continuous representation of the Heisenberg group. Note that if the view
of semiclassical approximation mentioned above is to have a general
enough validity, then it is necessary to be able to construct non-linear
phase spaces as well and the quotienting procedure given here shows that
it is possible.

A useful semiclassical approximation however, cannot be developed
without reference to dynamics. Construction of effective classical
dynamics induced from the quantum one required the choice of a section
of the bundle.  Once a section ({\em any section}) is chosen, one can
immediately define an effective Hamiltonian (and other effective
functions) on the classical phase space and a corresponding classical
dynamics. However, this associated classical dynamics will be a poor
approximation to the underlying quantum mechanics {\em unless} the
sections are further restricted.  We discussed the necessary conditions
on the section. The two main conditions are those of semiclassicality
and tangentiality.  However, when $\Gamma_{cl} \ne \Gamma$, one needs
additional condition namely, there should be sufficiently many quantum
motions which will project to curves in $\Gamma_{cl}$ (and not just in
$\Gamma$) and that these comprise a (local) section. 

The polymer state space brought out further features. To have a
non-trivial motion on $\Gamma_{cl}$, one needs to enlarge the set of
basic operators to include trigonometric operators with a set of
incommensurate $\lambda$'s, use states which are based on
(sub-)lattices generated by these $\lambda$'s {\em and} have the
Hamiltonian operator also depend on trigonometric operators with many
$\lambda$'s, except when the Hamiltonian operator has a additively
separated dependence on the trigonometric operators and the operator
$\hat{p}$. Furthermore, the semiclassicality criteria also needs to
be applied to the trigonometric operators such that the
uncertaintities for the operators labelled by $\lambda$ are computed
with states based on (sub-) lattice generated by the same label
$\lambda$. In effect, one chooses states based on the various
$\lambda-$ lattices and chooses a linear combination of these states
to form a (local) section. How to choose the set of incommensurate
$\lambda$'s is not clear at this stage.  A detailed construction of a
semiclassical approximation for LQC, following the steps discussed
here, needs to be done. The new Hamiltonian operator proposed in
\cite{APS} looks promising since it is self-adjoint and also
naturally connects triad labels in a specific irregular lattice. A
detailed analysis of the original LQC Hamiltonian of \cite{ABL}, with
coherent states based on a single $\lambda-$lattice is available in
\cite{Willis}.

There is an alternate way to develop a semiclassical approximation
\cite{EffectiveEquations}. This also uses the geometric view of
quantum mechanics and looks at the Hamiltonian flow on the quantum
phase space directly. For the special case of Hilbert space carrying
the weakly continuous representation of the Heisenberg group, one can
introduce the so-called {\em Hamburger momenta} variables to
introduce suitable (adapted to the bundle structure) coordinates on
the quantum state space. The exact quantum dynamical equations can
then be viewed as the Hamilton's equations of motion. Depending upon
the quantum Hamiltonian function, evolution of most of these
variables could decouple from those of a smaller (finite) set of
variables.  This smaller set of variables would then constitute a
classical approximation i.e. be thought of as classical degrees of
freedom. In effect, the evolution of the remaining `quantum degrees
of freedom', control how the uncertaintities in the classical degrees
of freedom evolve. Violation of semiclassicality can then be viewed
as coupling of the evolutions of the classical and the quantum
degrees of freedom. This method would be more useful to track when
the quantum evolution can exit the semiclassical bands signalling
break down of semiclassical approximation. So far this method has
been available in the context of Schrodinger quantization. Such an
explicit description of quantum evolution probably has to be done on
a case by case basis.

Clearly, within the restrictions provided by semiclassicality and
tangentiality, one can imagine different approximation schemes which will
systematically construct a better and better semi-classical
approximation. A natural way to phrase such a procedure is to formulate
it in terms of a family of sections. Thus one may begin with a section
as giving leading classical approximation and systematically change it
to improve the approximation. The usual perturbative approach can be
viewed as beginning with a section defined by ``free particle states''
and adding corrections to it get the new section closer to
tangentiality. This would be an interpretation of inclusion (or
computation) of quantum corrections.

One also encounters a situation wherein new degrees of freedom are
excited beyond a certain threshold scale. This will mean that merely
changing sections will {\em not} ensure tangentiality and
semiclassicality.  In the language of \cite{EffectiveEquations}, some of
the `quantum degrees of freedom' have to be thought of as new `classical
degrees of freedom'.  One has to include further `basic' operators and
use a new quotienting procedure.  Now one can repeat the analysis and it
is certainly conceivable that the true dynamics does satisfy the two
requirements with these additional variables defining sections.

Thus the geometrical view point shows that from a quantum perspective,
inadequacy of a classical approximation can arise in two ways --
inappropriate  choice of section which could be improved perturbatively
in some cases and/or inadequate choice of basic variables used in the
quotienting procedure.

The present analysis leaves out two important classes of systems: finite
dimensional constrained systems and field theories with or without
constraints. We hope to return to these in the future.

\begin{acknowledgments}
Part of this work was completed during a visit to IUCAA, Pune in May.
The warm hospitality is gratefully acknowledged. I would like to thank
Alok Laddha for discussions.
\end{acknowledgments}

\end{document}